\documentclass[manuscript]{aastex}
\usepackage{booktabs}
\usepackage{fixltx2e}

\shorttitle{Local Bubble Survey}
\shortauthors{Farhang et al.}

\begin{document}
\title{Probing the Local Bubble with Diffuse Interstellar Bands. III. The Northern hemisphere data and catalog}

\author{Amin Farhang\altaffilmark{1,2}, Habib G. Khosroshahi\altaffilmark{1}, Atefeh Javadi\altaffilmark{1}, Jacco Th. van Loon\altaffilmark{3}}

\email{a.farhang@ipm.ir}

\altaffiltext{1}{School of Astronomy, Institute for Research in Fundamental Sciences (IPM), P. O. Box 19395-5746, Tehran, Iran}
\altaffiltext{2}{Department of Physics, Sharif University of Technology, P. O. Box 11365-9161, Tehran, Iran}
\altaffiltext{3}{Astrophysics Group, Lennard-Jones Laboratories, Keele University, Staffordshire ST5 5BG, UK}


\def \kms {\rm{km~s^{-1}}} 
\def \hii {\mbox{H\,{\sc ii~}}}
\def \hi {\mbox{H\,{\sc i~}}}
\def \na {\mbox{Na\,{\sc i~}}}
\def \ca {\mbox{Ca\,{\sc ii~}}}
\def \do {\mbox{$\lambda 5780~$}}
\def \dt {\mbox{$\lambda 5797~$}}
\def \bv {\mbox{E\,{\textsubscript {B-V}~}}}


\begin{abstract}
We present a new high signal-to-noise (S/N) observations of the Diffuse Interstellar Bands (DIBs) in the Local Bubble and its surroundings. We observed 432 sightlines and obtain the equivalent widths of \do and \dt \AA~DIBs up to distance of $\sim$ 200 pc. All observations have been carried out by using Intermediate Dispersion Spectrograph (IDS) on 2.5 m Isaac Newton Telescope, during three years, to reach a minimum S/N ratio of $\sim$ 2000. All \do and \dt absorptions are presented in this paper and the observed values of interstellar parameter; $\lambda 5780$, $\lambda 5797$, \na D$_{1}$ and \na D$_{2}$ including the uncertainties are tabulated. 
\end{abstract}



\section{Introduction}

Sun is located within a hot bubble (known as the Local Bubble) with a low neutral gas density of $n(\rm{H}) \sim 0.01 $ cm$^{-3}$ \citep{Bohlin75,Weaver77} and purportedly hot temperature ($T \sim 10^{6}$ K, but see \cite{Welsh09}), estimated based on the distribution of diffuse soft X-ray background emission \citep{Snowden98}. According to recent observations of neutral gases (\mbox{Na\,{\sc i}}), ionized atoms (\mbox{Ca\,{\sc ii}}) and dust grains, the Local Bubble is depleted from these common species \citep{Vergely01, Lallement03, Welsh10}. Observation of nearby hot white dwarfs within the Local Bubble, failed to detect acceptable {O\,{\sc vi}} absorption level \citep{Oegerle00}. In addition, other observations of nearby white dwarfs and hot stars also failed to detect acceptable levels of highly ionized gas such as {C\,{\sc iv}} and {Si\,{\sc iv}} \citep{Bertin95,Holberg99} within the Local Bubble.

The 3D gas maps around the Local Bubble reveal that this cavity has a chimney like structure extended to a distance of $\sim$ 80 pc in the Galactic Plane and up to hundreds of pc into the Halo \citep{Vergely01, Lallement03, Welsh10}. Also the diffuse X-ray background emissions of the Local Bubble and its neighboring cavities, have been studied to find a relation between shape of the cavities and the soft X-ray data \citep{Puspitarini14}. While the origin of the Local Bubble is still debated, the most plausible model is that the Local Bubble has been created as a result of a number of successive supernovae in the nearby Sco$ - $Cen association \citep{Smith01}.

In recent years about $\sim 500$ narrow to broad interstellar absorption features have been observed which are known as Diffuse Interstellar Bands (DIBs) \citep{Herbig95}. The carrier of these heavy molecules are unknown, however, the DIB candidates could be among an infinite number of large carbon-based "organic" molecules \citep{Sarre06}. DIBs are used to study different condition of ISM, for instance some outstanding DIBs in the optical waveband are empirically known to trace the neutral phase of the interstellar medium (ISM). Also DIBs are correlated well with the color excess E\,{\textsubscript {B-V}}, the neutral hydrogen and the \na column densities \citep{Herbig93,Friedman11}.

DIBs react to the strength of the UV field of the local environment \citep{Cox06} and DIB strength varies as a function of UV radiation toward different sightlines \citep{Vos11}. Therefore, specific groups of UV resistant molecules, such as Polycyclic Aromatic Hydrocarbons (PAHs), fullerenes and carbon chains are commonly the more acceptable candidates for DIB carriers \citep{Herbig95}. DIB carriers have been ubiquitously detected everywhere, for instance, DIBs have been observed in the M31 and M33 \citep{Cordiner08a,Cordiner08b}, Magellanic Clouds \citep{vanLoon13} and beyond the Milky Way in SN host galaxies \citep{Cox08}. 

Since the temperature of the Local Bubble is purportedly high, the typical atoms and molecules could not survive in such an environments. In addition, as mentioned above, any attempts to detect other highly ionized atoms, or search for acceptable level of EUV emission have failed. Therefore we study the Local Bubble and its surroundings through DIBs in a survey over a 3 years period. We present the sightlines, spectrums, fitted profiles and the equivalent widths (EWs) for the northern hemisphere survey.

\section{OBSERVATIONS}
\label{observ}
All observations have been obtained with the Intermediate Dispersion Spectrograph (IDS) at the 2.5 m Isaac Newton Telescope (INT) at the Roque de Los Muchachos in La Palma. The IDS employs a long-slit spectrograph with two CCDs: The EEV10 CCD sensitive in blue, and the RED+2 which is sensitive in the red and both CCDs have $4096\times2048$ pixels. The spatial scale for the EEV10 is $ 0''.4$ and for the RED+2 is $ 0''.44 $ $/$pix, and the full unvignetted slit length is $ 3'.3 $.

For our observations we have used the 235 mm camera and H1800V IDS grating for an effective resolution of 0.31 \AA$/$pix. H1800V was chosen since it provides a high spectral resolution well matched to the typical width of DIBs. Also we have chosen 5800 \AA~as the central wavelength. A $ 1'.1 $ slit yielded spectra in the 5750$-$6040 \AA~region at spectral resolutions of $R\equiv \lambda / \Delta \lambda \sim2000$ (or a velocity resolution of $\Delta v = 150$ km s$^{-1}$). The DIB detection requires a high signal-to-noise (S/N) ratio of at least 100 but for detection of very weak absorptions, as like the one was detected by \cite{Cordiner06} towards the nearby star $\mu^{1}$Cru with \do DIB equivalent width (EW) of 4 m\AA, we need S/N of at least 2000. The seeing varied during these observing nights from 1\arcsec.1 to 1\arcsec.9.

To achieve a high quality data, we obtained more than 60 flat field frames with quartz lamp (exposure times of 13$-$15 sec), 15 arc frames (CuAr+CuNe) for wavelength calibration and a large number of bias frames every night. In order to achieve a minimum S/N $\sim$ 2000, we exposed each target 9$-$25 times, depending on their apparent magnitudes (see below). The data were processed using the {\sc CCDRED} data reduction package and the spectrum were extracted using the {\sc KPNOSLIT} package.

We have selected bright stars up to a distance of 200 pc. Targets were selected from the 3D \na D lines survey of \cite{Welsh10}, all with well-known distances from the Hipparcos satellite \citep{Perryman97}. Since the saturation level of the IDS detector was $\sim 64000$ counts, a small number of targets were rejected due to their extreme brightness ($V < 1.8$ mag). To increase the observing run efficiency, targets with $V > 7.2$ mag were also rejected from our target list to avoid very long integration times. Also to maximize the uniformity of the survey, in addition to the observation of hot stars (O, B types) we also observed some cooler stars (A, F, G \& K).

\section{Data analysis}
\label{ewmeas}
In DIB studies the shape of spectral profiles remain uncertain, since the carriers are unidentified. In addition, blending by that of other chemical species and possibility for blending with features from higher rotational levels of the same species \citep{Friedman11}, make DIB studies challenging. The equivalent width is defined as:

\begin{equation}
W = \int \frac{I_{0}(\lambda)-I(\lambda)}{I_{0}(\lambda)}~d\lambda
\label{eqew}
\end{equation}

Where, the $I_{0}$ and $I_{\lambda}$ are fluxes of the continuum and the spectral line respectively. With a Gaussian function one can approximate the \do and \dt lines absorptions \citep[e.g.,][]{Jacco09}. Therefore we obtain the line width in terms of the $\sigma$ value of the Gaussian distribution, and then calculate the Full-Width at Half Maximum (FWHM) as FWHM=$2(2~\ln2)^{1/2} \sigma = 2.355 \sigma$. Accordingly for producing a normalized spectrum, continuum fitting to the observed spectra was performed by fitting a nine-order Legendre polynomial.

For computing the statistical uncertainty for each observed DIB, we compute the standard deviation of residuals of the Gaussian fitting and summed in quadrature, and weighted by the Gaussian fit \citep{Jacco09,Vos11}. In DIB studies, the main source of error in the equivalent widths is the uncertainty in determining the real position of the continuum (systematic error), therefore the statistical error is always underestimated. For computing the systematic error, we fit three different continuum lines to local absorption region with $\pm 12$ \AA~range around the central DIB wavelengths (linear fit to the continuum, quadratic fit to the continuum and fit to DIB with linear continuum \citep{Kos13}). Accordingly, we set the intersection points of the DIB absorption and the continuum level \citep{Krelowski93}. Therefore, the systematic uncertainty is determined by the difference between the highest and lowest values of equivalent width among these three EWs.

For extracting the ISM absorption of cold stars, it is necessary to simulate the spectrum of cold stars. For constructing the synthetic spectrum of cold stars, we calculate the atmosphere models with the ATLAS9 code \citep{Kurucz92}. We use grids of ATLAS9 model atmospheres from \cite{Castelli03}\footnote{http://wwwuser.oats.inaf.it/castelli/} with the new Opacity Distribution Functions (ODFs) for several metallicities \citep{Castelli97}. We obtained the effective temperature of the observed star and the surface gravity $\log g$ from \cite{Cayrel96, Varenne99, Reddy03, Prugniel07,Huang10,Prugniel11} and \cite{Koleva12}. Also we determine target metallicity from \cite{Heacox79, Abt02, Royer02, Royer07} and \cite{Schroder09}. After modeling the stellar atmosphere we generate the synthetic stellar spectrum with the Linux port of the SYNTHE suite codes \citep{Sbordone04, Sbordone05, Kurucz05}. Atomic and molecular data were taken from the database on Kurucz website\footnote{http://kurucz.harvard.edu} \citep{Kurucz05}. Also for considering the broadening caused by rotational velocity, we use the rotational velocity from \cite{Abt02, Royer02, Royer07} and \cite{Schroder09}. For those stars which the rotational velocity was not reported, we use $v \sin i=15-20$ km s$^{-1}$. In our calculations we included all the atomic and molecular lines with empirically determined atomic constants plus all the diatomic molecular lines (CH, NH, CN, MgH, SiH, SiO, H$ _{2} $, C$_{2}$ and CO) except the TiO molecule.

The A-type stars have two main sources of contaminations \mbox{Fe\,{\sc ii}} (at 5780.13 \AA~and 5780.37 \AA) and \mbox{Fe\,{\sc ii}} (at 5783.63 \AA) at $\lambda5780$ position. As well as there is no source of contamination at $\lambda5797$. For correct these contaminations for each A-type stars that the metallicity have not been reported, we produce all possible synthetic spectra and compare them with the observed spectra. The surface gravity of A-type stars varies between 3.5 to 4.2 \citep{Gray92}, therefore we choose a constant $\log g = 4$ in all of our atmosphere model. From A0 to A9 spectral types, according to \cite{Theodossiou91} report, we select a constant temperature for each I, II$-$III and IV$-$V luminosity classes. Also the A-type stars have different metallicities from 0 to $ -2 $ \citep{Beers01}. Accordingly, for a given A-type subdivision (e.g., A0) and luminosity class (e.g., IV), we produce three different atmosphere models with [Fe/H] $= 0, -0.5, -1.5$, and compare with the observed spectrum to choose the best model. Also we consider the effect of rotational velocity convolved with the instrument dispersion. But the rotational velocity of our observed A-type stars are very high ($\sim 200$ km s$^{-1}$) \citep{Hoffleit95}, thus when convolved with our instrument dispersion, the absorption lines are widened. Therefore their stellar features impact on the DIB absorption will be limited. Then we subtract the synthetic spectrum from the observed absorption (containing both stellar and interstellar absorption features). The obtained residual predominantly consists of the interstellar absorption \citep{Montes95a,Montes95b}. Then by fitting a Gaussian function to this residual the equivalent width of the DIB will be obtain.

The spectrum shape of cool F,G, and K type stars, is similar to a two prong fork. This absorptions are caused by the presence of \mbox{Fe\,{\sc ii}}, \mbox{Mn\,{\sc i}}, \mbox{Si\,{\sc i}} (all near 5780.1 \AA) and \mbox{Cr\,{\sc i}} (5781.1 \AA). Therefore for producing the synthetic spectrum of these cool stars, we consider a Gaussian profile for \do absorption lines:

\begin{equation}
D(\lambda) = a~\exp \left(  -\frac{(\lambda - b)^2}{2\sigma}  \right) + c
\label{eqd}
\end{equation}

After confirming the average \do DIB profile in an iterative procedure, according to Eq.~\ref{eqd}, we change the $ a $ (peak intensity), $ b $ (peak center) and $\sigma$ (peak with) for each wavelength ($\lambda$) to produce a new spectrum. Then add this spectrum with the corresponding synthetic spectrum, and in each iteration according to Eq.~\ref{eqchi}, calculate the difference of the re-produced spectrum with the real observed absorption ($\chi^{2}$). Eventually, the best DIB profile estimation is one with the smallest $\chi^{2}$ value.

\begin{equation}
\chi^{2} = \sum_{i=1}^{N}\left(\frac{\left( F(\lambda_{i}) - D(\lambda_{i}) \right)^{2}}{F_{{\rm err}}(\lambda_{i})^{2}}  \right)
\label{eqchi}
\end{equation}

We show in plot the observed spectra and their best Gaussian fit to \do and \dt features. The red fits are those with acceptable DIB absorption and the blue fits are those that we are not confident about DIB character. Also the table shows the first 10 samples of \do equivalent width and its uncertainty, the \dt equivalent width and its uncertainties, the \na D$_{2}$ and \na D$_{1}$ and their uncertainties  and the target distances. Last three columns are the $\lambda5780$, \dt and \na flags, which 1 means the quality of measurements are acceptable and 0 means unacceptable values. The full tables and plots as soon as will be released.

\acknowledgments
We wish to thank the Iranian National Observatory (INO) and School of Astronomy at IPM for facilitating and supporting this project. The observing time allocated to this project was provided by the INO. We also wish to thank the ING staff for their support. Some of the research visits related to this project have been supported by the Royal Society International Exchange Scheme. Furthermore, we thank the referee for a thorough reading of the manuscript and useful comments and suggestions.



\clearpage
\begin{figure*}
	\includegraphics[scale=0.8]{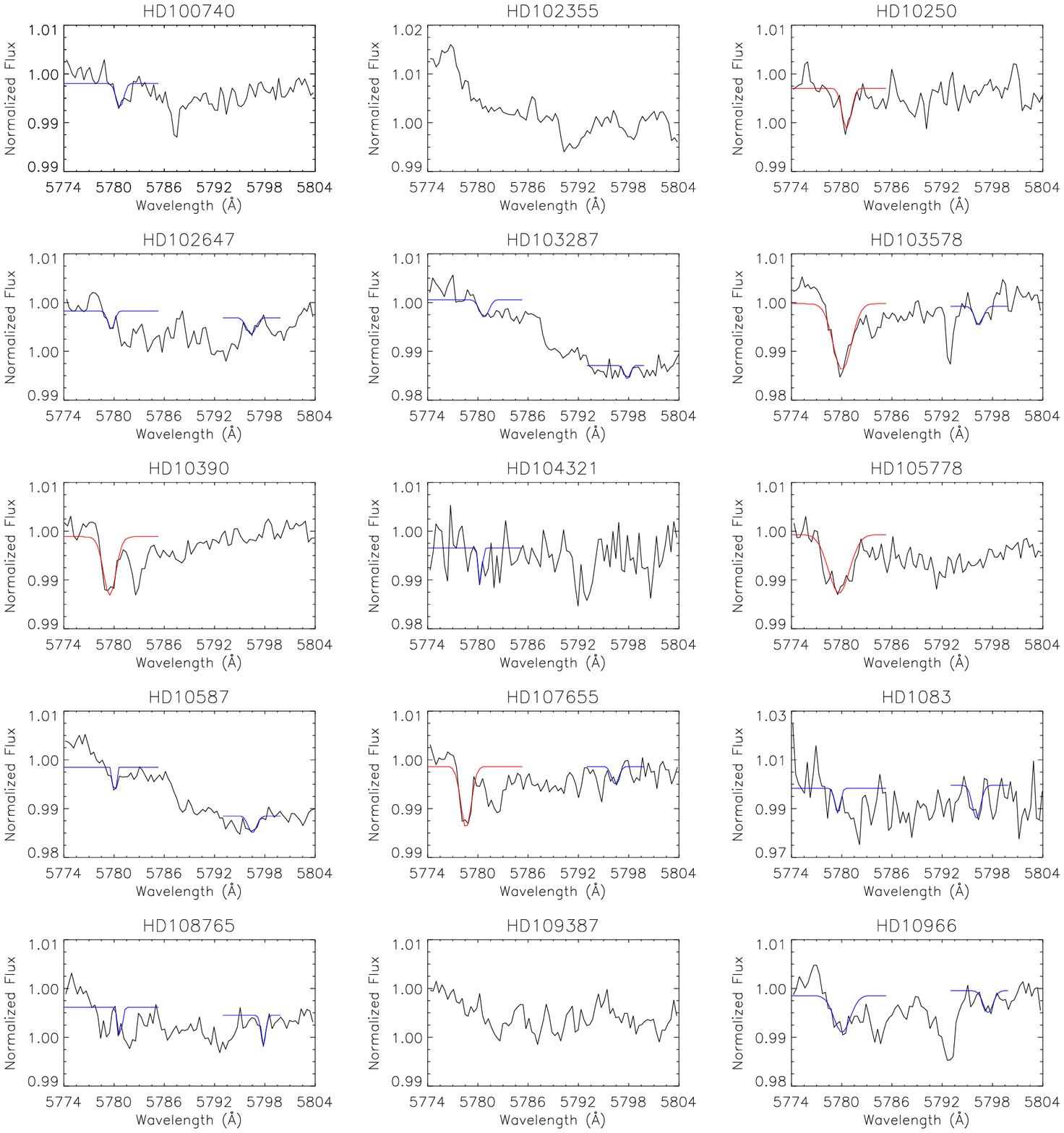}
	\caption{The observed spectrum of nearby stars up to distance of $\sim$ 200 pc. The Gaussian fitted lines are divided to two different categories; the red fitted profiles shows acceptable DIBs and the blue lines represents the uncertain DIB features.}
	\label{f1}
\end{figure*}


\clearpage
\begin{deluxetable}{lcccccccccccccc}
	\tablewidth{0pt}
	\tabletypesize{\scriptsize}
\rotate

	\tablecaption{The EW measurments of $\lambda5780$, \dt and \na absorptions in the northern hemisphere up to distance of $\sim$ 200 pc.}

	\tablehead{
		\colhead{Name}           & \colhead{glon}      &
		\colhead{glat}          & \colhead{dis}      &
		\colhead{$\lambda5780$}          & \colhead{err $\lambda5780$}   &
		\colhead{$\lambda5797$}       & \colhead{err $\lambda5797$}   &
		\colhead{\na D$_{2}$}       & \colhead{err \na D$_{2}$}      &
		\colhead{\na D$_{1}$}          & \colhead{err \na D$_{1}$}      &
		\colhead{f1}             & \colhead{f2}       &
		\colhead{f3}}
	\startdata

   HD10250  &   127.2  &     8.2  &     86  &    7.47  &    1.95  &    0.00  &    0.00  &    0.00  &    0.00  &    0.00  &    0.00  &     1  &     0  &     0   \\
   HD10390  &   134.5  &   -26.5  &     79  &   16.30  &    2.97  &    0.00  &    0.00  &    0.00  &    0.00  &    0.00  &    0.00  &     1  &     0  &     0   \\
   HD10587  &   130.2  &    -5.0  &    172  &    3.33  &    0.93  &    4.64  &    1.25  &    0.00  &    0.00  &    0.00  &    0.00  &     0  &     0  &     0   \\
    HD1083  &   113.1  &   -34.9  &    126  &    7.75  &    2.74  &   16.42  &    3.14  &    0.00  &    0.00  &    0.00  &    0.00  &     0  &     0  &     0   \\
   HD11291  &   132.6  &   -10.9  &    151  &    0.00  &    0.00  &    0.00  &    0.00  &   46.34  &   12.36  &   69.17  &   11.85  &     1  &     1  &     1   \\
   HD11335  &   132.6  &   -10.2  &    152  &    3.53  &    2.10  &    0.00  &    0.00  &   65.42  &   11.72  &   56.51  &   10.97  &     0  &     0  &     1   \\
   HD11415  &   129.8  &     1.6  &    136  &   22.59  &    3.58  &    1.63  &    0.67  &   55.26  &    9.18  &   33.96  &    8.52  &     1  &     0  &     1   \\
   HD11946  &   130.2  &     2.7  &     79  &    6.46  &    1.70  &    0.00  &    0.00  &    0.00  &    0.00  &    0.00  &    0.00  &     1  &     1  &     0   \\
   HD12216  &   128.4  &    10.3  &     50  &   16.73  &    2.81  &    2.17  &    0.76  &  161.15  &   16.00  &  251.97  &   16.19  &     1  &     0  &     0   \\
    HD1280  &   115.6  &   -23.7  &     78  &   27.03  &    4.63  &    6.95  &    1.15  &  224.64  &   18.10  &  275.48  &   18.60  &     1  &     1  &     0   \\

	\enddata
	\tablenotetext{Notes}{Column 1: Star name (the Henry-Draper (HD) number), Column 2: galactic longitude (degree), Column 3: galactic latitude (degree), Column 4: distance (pc), Column 5: \do EW (m\AA), Column 6: \do EW uncertainty (m\AA), Column 7: \dt EW (m\AA), Column 8: \dt EW uncertainty (m\AA), Column 9: \na D$_{2}$ EW (m\AA), Column 10: \na D$_{2}$ EW uncertainty (m\AA), Column 11: \na D$_{1}$ EW (m\AA), Column 12: \na D$_{1}$ EW uncertainty (m\AA), Column 13: \do quality flag; 1 $=$ acceptable and 0 $=$ unacceptable value, Column 14: \dt quality flag; 1 $=$ acceptable and 0 $=$ unacceptable value, Column 13: \na quality flag; 1 $=$ acceptable and 0 $=$ unacceptable value.}


\end{deluxetable}

\end{document}